\newif\ifclns \clnsfalse

\clnstrue

\def\lam    { \Lambda \overline{\Lambda} }
\def\pp     { p \overline p }
\def\sig    { \Sigma^o \overline{\Sigma}{}^o} 
\def\sigo   { \Sigma^o }
\def\sigob  { \overline{\Sigma}{}^o }
\def\sors   { (\Sigma^o/\sigob) }
\def\lamsig { \Lambda \sigob }
\def\siglam { \Sigma^o \overline{\Lambda} }

\def\eeraee { e^+ e^- \rightarrow e^+ e^- }

\def\eell   { \eeraee \lam }
\def\eess   { \eeraee \sig }
\def\eellss { \eeraee (\Lambda/\sigo) (\overline{\Lambda}/\sigob)}
\def\eelssl { \eeraee (\lamsig/\siglam) }

\def\ggra   { \gamma \gamma \rightarrow }
\def\gglam  { \ggra \lam }
\def\ggll   { \ggra \lam }
\def\gglssl { \ggra (\lamsig/\siglam) }
\def\ggss   { \ggra \sig }
\def\ggpp   { \ggra \pp }

\def\mll    { m_{\lam} }
\def\mm     { m_{p \pi^-}, m_{\overline{p} \pi^+} }
\def\lumi   { 3.5 \, {\rm fb}^{-1} }

\def\pacdefs {12.38.-t, 12.38.Qk, 12.39.-x, 13.60.Rj, 13.60.-r, 14.20.Jn}

\def\abdef {
Using the CLEO detector at the Cornell $e^+e^-$ storage ring, CESR,
we study the two-photon production of $\lam$,
making the first observation of $\ggll$. 
We present the cross-section for $\gglam$ 
as a function of the $\gamma \gamma$ center of mass energy and compare it to
that predicted by the quark-diquark model. }

\ifclns
\def\setfigsize{ \epsfysize=10cm }
\else
\def\setfigsize{ \epsfysize=7.5cm }
\fi

\ifclns
\documentstyle[aps,prl,epsf,floats,preprint]{revtex}
\else
\documentstyle[aps,prl,epsf,floats]{revtex}
\fi

\begin{document}


\ifclns
\preprint{\tighten\vbox{\hbox{CLNS 96/1448 \hfill}
                        \hbox{CLEO 96-19   \hfill}
                        \hbox{January 17, 1997 \hfill}}}

\fi

\title{$\Lambda \overline{\Lambda}$ Production in 
   Two-photon Interactions at CLEO}

\ifclns

\maketitle

\begin{center}
CLEO Collaboration
\end{center}

\tighten

\begin{abstract}
  
  \abdef
  
\end{abstract}

\pacs{ \pacdefs}

\clearpage

\begin{center}
S.~Anderson,$^{1}$ Y.~Kubota,$^{1}$ M.~Lattery,$^{1}$
J.~J.~O'Neill,$^{1}$ S.~Patton,$^{1}$ R.~Poling,$^{1}$
T.~Riehle,$^{1}$ V.~Savinov,$^{1}$ A.~Smith,$^{1}$
M.~S.~Alam,$^{2}$ S.~B.~Athar,$^{2}$
% ...
\end{center}
 
\small
\begin{center}
$^{1}${University of Minnesota, Minneapolis, Minnesota 55455}\\
$^{2}${State University of New York at Albany, Albany, New York 12222}\\
% ...
\end{center}

\fi

\ifclns
   \thispagestyle{empty}
   \clearpage
\fi

%
%

Two-photon interactions are a useful tool for the study of the strong 
interaction.  
At CLEO we use two-photon interactions to test calculations 
of strong processes as well as the understanding of hadron structure. 
CLEO has previously measured the cross-section for $\ggpp$ \cite{Ong}.  
Extending this analysis, in this paper we report on the study of
$\ggll$.

Using the Brodsky-Lepage hard-scattering approach \cite{bandl}, 
predictions have been made for the two-photon production of baryons.  
The CLEO measurement of the $\ggpp$ 
cross-section was inconsistent with the prediction of a pure quark 
model \cite{Farrar} at energies available to CLEO, 
but was consistent with 
the prediction of the quark-diquark model \cite{Anselmino,Kroll}.  
We compare the measured $\ggll$ cross-section
to that predicted by these models.

CLEO II is a general purpose detector\cite{cleo} using
the $e^+ e^-$ storage ring, CESR\cite{cesr}, 
operating at $\sqrt{s}\sim \! 10.6$ GeV.  
CLEO II contains three concentric wire chambers
that detect charged particles over 95\% of the solid angle.
Particle identification is performed using specific ionization energy loss
(dE/dx) in the outer wire chamber. 
A superconducting solenoid provides a magnetic field of 1.5 T, giving
a momentum resolution of  
$\sigma_p/p \approx 0.5 \%$ for $p = 1$ GeV.  
Outside of the wire chambers and a time of flight system, but inside 
the solenoid, is a CsI electromagnetic calorimeter, 
consisting of 7800 crystals arranged
as two endcaps and a barrel.  
For a 100 MeV electromagnetic shower in the barrel,
the calorimeter achieves an energy resolution of
$\sigma_E/E \approx 4\%$.

Kinematics of two-photon events are strongly influenced
by the fact that the initial state photons are
approximately real and tend to have a large fraction
of their momenta along
the beam line.
A typical $|q^2|$ of the photons is $20 \, {\rm MeV}^2$,
where $q$ is the photon four-momentum. 
Consequently, the two-photon axis is approximately
the beam axis,
and the electron and positron rarely 
have enough transverse momentum to be observed.
The two photons have rather unequal energies, 
causing the $\lam$ center of mass to be boosted along
the beam axis.  
As the available energy in the
$\Lambda$ decay is small, and the $\ggll$ 
cross-section is
peaked near the $\lam$ threshold,
the decay products, 
$p \pi^- \overline p \pi^+$, 
usually have relatively low 
transverse momentum.
We select those events
in which all four hadronic tracks are observed in CLEO.  

In our analysis of $\lumi$ of data, 
we use the following selection criteria to minimize background.
We select 4 track events in which the charge sum is zero.
We require the candidate proton and antiproton
to have dE/dx measurements consistent 
with that of a proton. 
We require that the event energy, using these particle
assignments, is less than 6.0 GeV and
that the transverse component of the vector sum of the track momenta
is less than 0.6 GeV/c.  
We veto events in which the candidate 
$\Lambda$ or $\overline{\Lambda}$ vertex is at the radius of the
beam pipe.
We also place a requirement on the transverse impact parameters 
of the
reconstructed $\Lambda$ and $\overline{\Lambda}$ with respect 
to the transverse beam position;
their root sum square must be less than 1.0 cm.
Finally, cross-section predictions\cite{Anselmino,Schweiger_per} 
have been made for $|\cos{\theta^*}| < 0.6$,
where $\theta^*$ is the angle between the $\Lambda$ momentum 
and the two-photon axis in the two-photon center of mass frame.  
In order to compare with theory and with $\ggpp$ measurements,
we impose the same requirement on the data.
As the acceptance of the detector decreases quickly beyond
$|\cos{\theta^*}| = 0.6$,
this requirement does not significantly affect the event yield.
After applying these selection criteria, 
there is a clear enhancement in
$(m_{p \pi^-}, m_{\overline{p} \pi^+})$ plane
at the $(m_{\Lambda},m_{\overline{\Lambda}})$ point.
    
To verify that the reconstructed particles are predominantly
$\Lambda$'s and $\overline{\Lambda}$'s produced in
two-photon interactions, 
a number of data and Monte Carlo distributions have been
compared, including event energy, decay distance,
$\Lambda$ momentum angular distribution, proper decay length, 
acoplanarity, acolinearity,
proton momentum, and pion momentum.
In all cases there is good agreement between the data
and the expected distributions.  

The signal and background two-photon Monte Carlo events were
generated using a program based on the BGMS formalism\cite{Budnev}. 
The simulation 
of the transport and decay
of the final state particles through the CLEO detector
is performed by a GEANT-based detector simulator\cite{geant}.  

\begin{figure}
   \centering \leavevmode
   \setfigsize
   \epsfbox{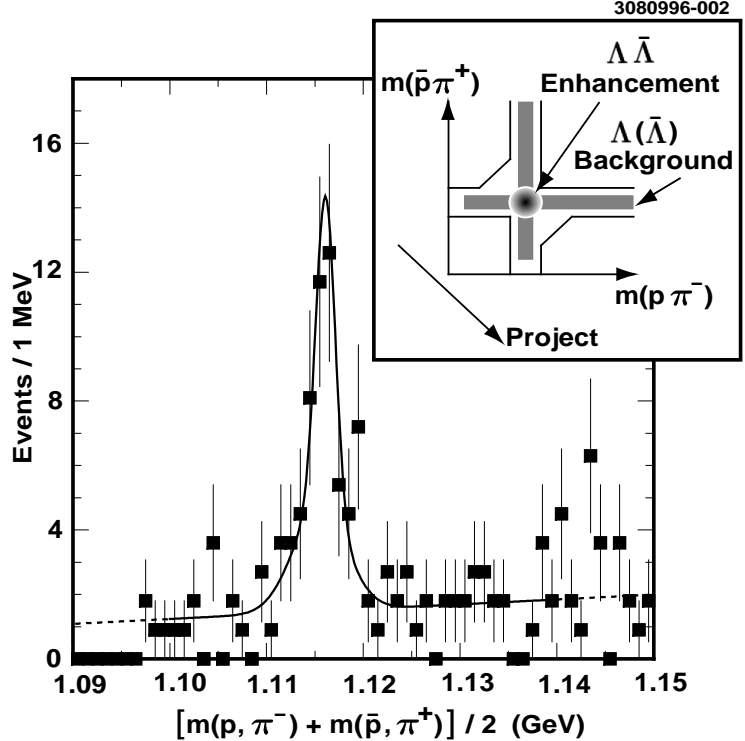}
   \caption{$\lam$ mass average distribution found in data.  
     Insert depicts mass average projection technique.}
   \label{fig_rot_mass}
\end{figure}

We use the $\lam$ mass average, 
$(m_{p \pi^-} + m_{\overline p \pi^+})/ 2$, to measure
the number of signal events\cite{ww}.
Viewed geometrically,
the mass average rotates the $(m_{p \pi^-}, m_{\overline p \pi^+})$
plane by $-\pi/4$ and then scales the projected value by $1/ \sqrt{2}$;
see Figure \ref{fig_rot_mass}.
The advantage of this approach is that it naturally
maps backgrounds from
$\Lambda$-fake, fake-$\overline{\Lambda}$, and fake-fake into
smooth backgrounds in the mass average plot which
are then easier to subtract when fitting.

We reduce the fake-fake 
background by making a geometric selection in the $(\mm)$
plane before the projection.  
We require that events are within 6 MeV,
nearly $4$ times the $\Lambda$ mass resolution,
of the $\Lambda$ mass for either axis.
This simple cross geometry would underestimate fake-fake background
near the $\lam$ enhancement.  
To compensate for this, we extend our 
geometric criteria at the intersection of the cross so that the
area along the projected direction is a constant.
This approach is valid as the fake-fake
background does not vary significantly near the $\lam$ enhancement.

Using a signal shape fit to the 
mass average distribution of the Monte Carlo 
combined with a linear background, 
we measure $51.0 \pm 8.6$ events in data.
The fit and data are displayed in
Figure \ref{fig_rot_mass}.

Due to reduced sensitivity to other channels and
the steep $W$
dependence of two-photon production, 
the dominant source of feeddown into
the observed signal comes from the two-photon
production of $\sig$, $\lamsig$ or $\siglam$,  
where $W$ is the two-photon center of mass energy.
At this point we have not used final state photon information to
distinguish between the four possible final states 
$(\Lambda/\sigo) (\overline{\Lambda}/\sigob)$
for which we use this parenthetical notation 
to indicate alternative processes.
 
In order to measure the cross-section
we apply the projection
technique to the selected data and Monte Carlo events 
with the addition that we
bin in $\mll$, the
effective mass of the $\Lambda$ and $\overline \Lambda$.
If the source of the signal is $\ggll$, then
$\mll = W$.
We fit to the background excluding the
signal region and subtract this from
the number of events within the signal region,
which is within about 3 times the mass resolution
of the $\Lambda$ mass.
The number of events summed over all bins is constrained to be 51.0,
the total number of events measured.
We have estimated the systematic uncertainty associated with binning,
selection criteria, and background shape.  
The non-negligible sources of
uncertainty are associated with
triggering, $13\%$, tracking, $14\%$,  
and event selection, $14\%$.
Assuming that these are independent, 
gives a quadrature sum of $24\%$,
which is conservative in this case.

We find the $\eellss$ cross-section in each bin of $\mll$
by correcting the observed $\lam$ yield in that bin
by the efficiency obtained from the Monte Carlo simulation.
Summing these we find the total
$\eellss$ cross-section for $|\cos \theta^*| < 0.6$
to be $2.0 \pm 0.5 \pm 0.5 \, {\rm pb}$.
The first error is statistical, dominated by the statistics
of the first bin, the second is the $24 \%$ systematic
uncertainty discussed above.
This corresponds to an overall efficiency of $1.8 \%$.

To correctly extract a cross-section, 
the contamination from feeddown into the observed signal 
must be removed.
As the statistics are limited, we do not use the
mass average technique, but instead we 
search for either a $\sors$ in the events 
that pass all $\lam$ selection criteria described above
and that lie within a $6 \, \rm{MeV}$ radius of the 
point $(m_{\Lambda}, m_{\overline{\Lambda}})$ 
in the $(m_{p \pi^-}, m_{\overline p \pi^+})$ plane.
To search for $\sors$ 
we combine each $\Lambda$ or $\overline{\Lambda}$ with
selected photon candidates in the event,
using the notation $\sors$ to indicate 
either a $\sigo$ or $\sigob$.

We only consider photon candidates in the crystal
barrel.  
The energy associated with the photon candidate must be
within the range 40 MeV to 180 MeV.
Each candidate photon must not be matched with an observed charged track, 
and we apply the stricter requirement that 
the cosine of the angle between 
the candidate photon
and the shower matched with the anti-proton track
must be less than $0.9$.
To reduce background from hadronic interactions, we require that 
the ratio of energy deposited in the central 9 crystals to
that in the central 25 crystals must be $> 0.9$.

\begin{figure}
   \centering \leavevmode
   \setfigsize
   \epsfbox{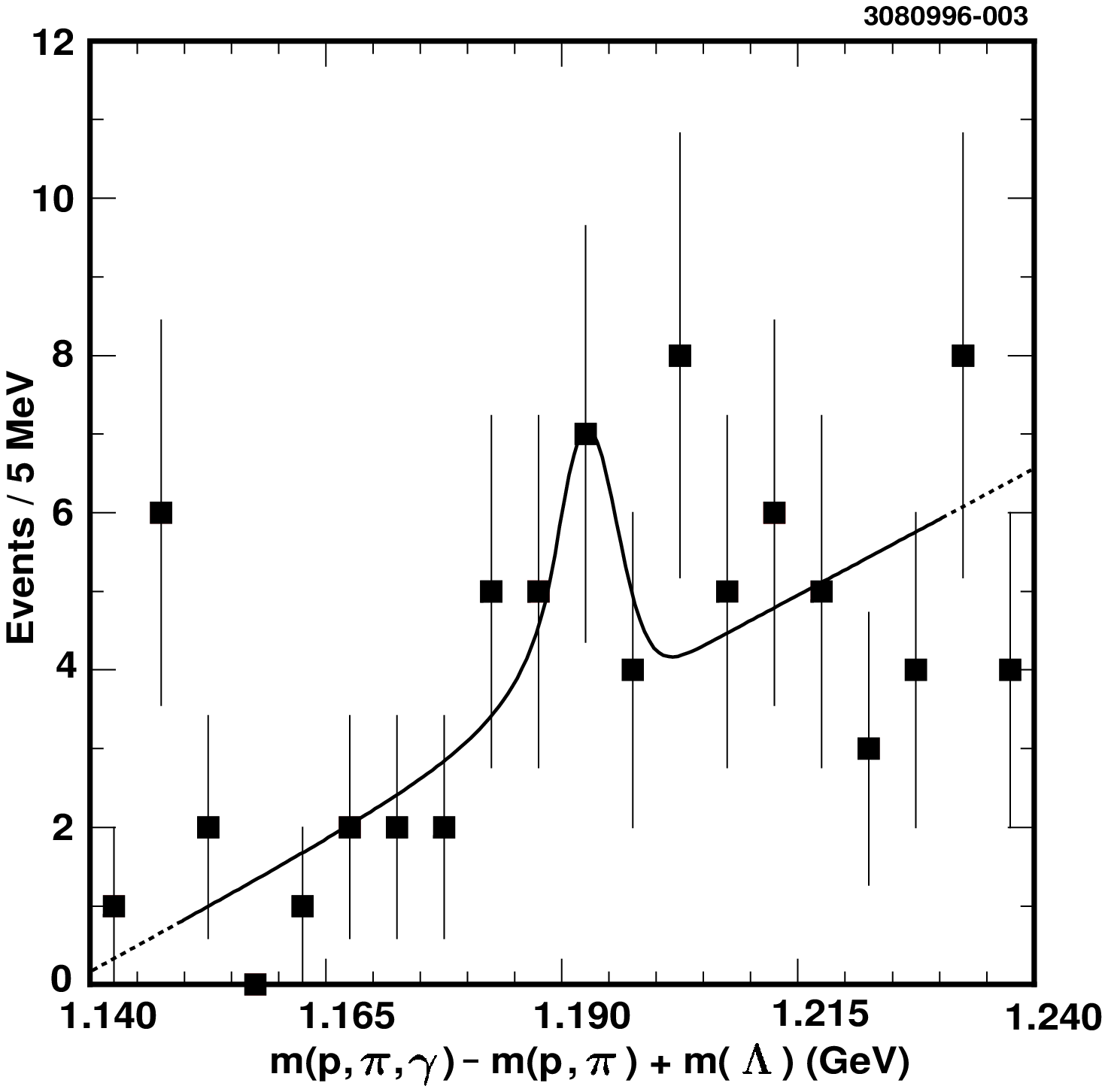}
   \caption{$\sors$ mass distribution found in data.}
   \label{fig_sig_fit}
\end{figure}

For each $\sors$ we construct   
$m_{p\pi\gamma} - m_{p\pi} + m_{\Lambda}$
which has better resolution than $m_{p\pi\gamma}$. 
We use a signal shape fit to the Monte Carlo distribution
combined with a linear background 
to fit the data. From the distribution in Figure \ref{fig_sig_fit} 
we measure the number of $\sors$ to be
$7.5\pm5.6$.  
As the statistical uncertainty associated with 
this measurement is very large, the systematic uncertainty
is not significant.
Although consistent with zero, this value will be used to estimate
feeddown into the $\ggll$ measurement.

Due to the low statistics of the $\sors$ measurement, 
we can not determine the ratio of 
$\gglssl$ to $\ggss$. 
We assume that the processes $\gglssl$ and $\ggss$ each contribute
half.
We assign the difference between the number given by the 
above mixture of processes
and that using the assumption that all 
observed $\sors$ events were 
produced by $\ggss$
as the systematic uncertainty.
Given the above ratio of contributions, 
we estimate the number of contamination events by multiplying
the observed number of $\sors$ by $4/3$ and by the
ratio of the $\lam$ detection efficiency to the $\sors$ detection
efficiency in $\eess$ events.  We have used the fact that
the efficiency for finding a $\sors$ in $\eess$ events
is approximately twice that for finding $\sors$ in $\eelssl$ events.
The estimated number of non-$\lam$
contamination events is $11 \pm 8 \pm 4$, 
giving a contamination correction scale factor of
$[1 - (11 \pm 8 \pm 4)/51.0] = 0.78 \pm 0.16 \pm 0.08$.
Applying this factor to the cross-section we extract an
exclusive $\eell$ cross-section of $1.6 \pm 0.6 \pm 0.4$ pb
for $|\cos \theta^*| < 0.6$.

To calculate the $\ggll$ cross-section 
we scale the measured signal using,

\begin{eqnarray}
  \sigma_{bin}^{data}
  &\approx&  { n^{data}_{bin}/L_{data}  \over
    n^{MC}_{bin}/L_{MC} }
  \, \sigma_{bin}^{MC} \, , 
  \label{eq_critass_last}
\end{eqnarray} 

\noindent to account for photon flux and efficiency,
in each $\mll$ bin.  
We correct the cross-section using
our estimate of the the $\sors$ contamination.
The $\mll$ distribution observed in data 
is a good model for the $\mll$ distributions of the 
$\sig$ and $(\lamsig/\siglam)$ contamination.  
Consequently, we can apply the contamination correction scale
factor bin by bin.  
The results are shown in Table \ref{tab_2pho_cor_cs}.
An additional systematic
uncertainty associated with the uncertainty
of the feeddown $\mll$ distribution is included.  

The predicted $\ggll$ cross-sections
appear to disagree with this measurement.
Due to the failure of the of pure-quark calculation 
to accurately predict the cross-section for $\ggpp$ 
at values of $W$ 
that we probe,
we do not anticipate that it can accurately
predict the cross-section for $\ggll$ \cite{Farrar}.
However, 
the quark-diquark model  
is constructed to predict
the cross-section in this 
energy regime.
This model includes nonperturbative effects 
through the use of the diquark, 
a $q q$ bound state within the baryon.  
The original calculations were performed
using only scalar diquarks \cite{Anselmino}.
More recent calculations include both
scalar and vector diquarks \cite{Kroll,Schweiger_per}.
In the energy regime near threshold,
the quark-diquark model is also expected to fail. 

\begin{figure}
   \centering \leavevmode
   \setfigsize
   \epsfbox{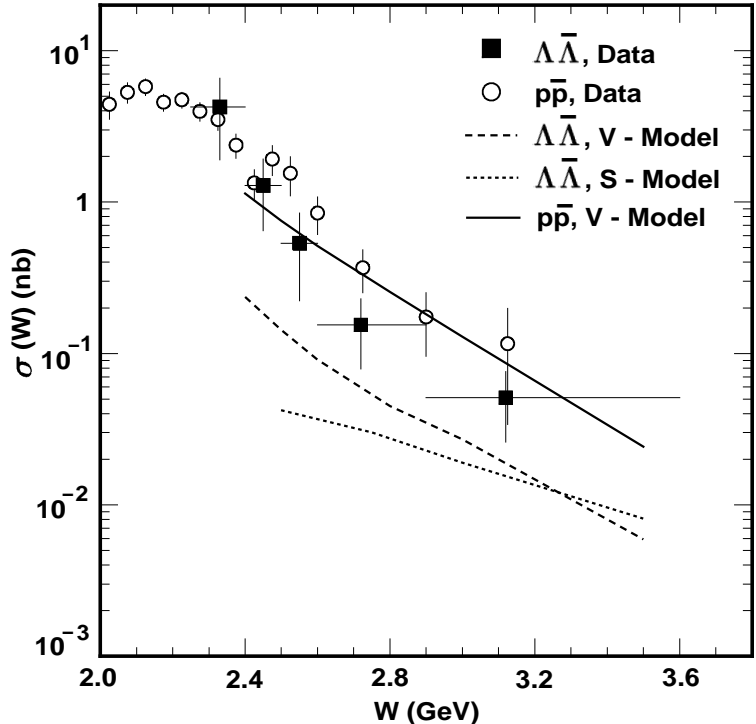}
   \caption{
        $\sigma_{\ggll}(W)$, 
        $\sigma_{\ggpp}(W)$
        for $|\cos{\theta^*}| < 0.6$. 
        Vertical error-bars include
        systematic uncertainties.  
        Horizontal markings indicate bin width.  
        S-model identifies the scalar quark-diquark model 
        \protect\cite{Anselmino}.
        V-model identifies the vector quark-diquark model
        \protect\cite{Kroll,Schweiger_per}.  
        }
   \label{fig_all_2pho_cs}
\end{figure}

The extracted exclusive $\ggll$ cross-section,
the previously measured $\ggpp$ cross-section, 
and the predictions of the model are
displayed in Figure \ref{fig_all_2pho_cs}
as a function of $W$ for $|\cos \theta^*| < 0.6$.
We place the horizontal 
location of the cross-section
data points at the weighted mean of $W$ in 
the bin based on a $\sim 1/W^{12}$ distribution.
We do not display the predictions of 
the pure-quark calculation, 
which are much
smaller than the quark-diquark predictions
for both $\ggpp$ and $\ggll$.
The unexpected result is that
the production of $\ggll$ appears to be consistently
larger than the prediction of the quark-diquark model.
In the three bins above 2.5 GeV 
the vector quark-diquark model predicts that we should 
observe $\sim 10$ events,
but in data we observe 32 events.

\begin{table}[tbh]
\ifclns
\begin{center}
\begin{minipage}{4.3in}
\fi
   \caption{Two-photon cross-section 
     $\ggll$ for $|\cos{\theta^*}| < 0.6$ }
   \begin{tabular}{c c}
      $\mll$ [GeV]  & $\sigma_{\ggll}$ [nb]  
      \\  \hline
      $2.25-2.4$ &  
         $4.2   \pm 1.7  \pm 1.6  $     \\
      $2.4-2.5$  & 
         $1.3   \pm 0.5  \pm 0.4  $     \\
      $2.5-2.6$  &  
         $0.54   \pm 0.27  \pm 0.16  $  \\
      $2.6-2.9$  & 
         $0.15  \pm 0.06 \pm 0.04 $     \\
      $2.9-3.6$  &
        $0.051  \pm 0.019 \pm 0.017 $     
      \\  
   \end{tabular}
   \label{tab_2pho_cor_cs}
\ifclns
\end{minipage}
\end{center}
\fi
\end{table}

In this paper we presented the first observation of $\ggll$.
We measured the $\eell$ cross-section,
and the $\ggll$ cross-section as a function of $\mll$,
each for $|\cos \theta^* | < 0.6$.
The measured $\ggll$ cross-section appears to be larger
than that predicted by either the quark-diquark model 
or the pure-quark calculation over the observed range of $W$.

We thank  P. Kroll and W. Schweiger
for access to their calculations.
We gratefully acknowledge the effort of the CESR staff in providing us with
excellent luminosity and running conditions.
This work was supported by 
the National Science Foundation, 
the U.S. Department of Energy, 
the Heisenberg Foundation,
the Alexander von Humboldt Stiftung,
the Natural Sciences and Engineering Research Council of Canada,
and the A. P. Sloan Foundation.

\end{document}